# Towards a Reflective PICOSEC detector?

**Amos Breskin**

*Department of Astrophysics and Particle Physics*
*Weizmann Institute of Science*
*Rehovot, Israel*
*E-mail*: amos.breskin@weizmann.ac.il

ABSTRACT: PICOSEC is an ultrafast particle-detector concept, combining a photocathode-coated Cherenkov radiator coupled to a gas-avalanche multiplier. Particle-induced Cherenkov photons create photoelectrons emitted from an ultrathin semitransparent photocathode; they are multiplied and detected in fast gas-avalanche mode. In parallel to the constant progress made in the PICOSEC technique, we propose different detector configurations and operation modes with the aim of enhancing robustness and performance. They incorporate thick reflective photocathodes deposited on the readout electrodes of various types of avalanche multipliers. Some of these Reflective-PICOSEC detectors operate at mbar gas pressures.

KEYWORDS: Electron multipliers (gas), Timing detectors, Photon detectors for UV, visible and IR photons (gas).

## 1. Introduction

Particle physics experiments at future high-luminosity accelerators pose ever-growing demands on radiation detectors [1]. Besides necessary advances in tracking, calorimetry, particle identification and other techniques - precise timing, in the few tens of picosecond range, is required to cope with the expected outstandingly high particle fluences.

Decades of important developments have been made in fast-timing gas-avalanche Resistive Plate Chambers (RPC), vastly deployed in current particle-physics experiments [2]. Their efficiency and timing properties are dictated by the statistics of energy deposition within very narrow gas gaps and particle flux. RPCs have yielded time resolutions in the sub-100ps range in large area single-layer RPC-detector systems. Multi-layer RPC (MRPC) detectors have yielded record resolutions of ~20ps - though in a 24-layer 18cm$^2$ area MRPC [3]. Constant efforts have been conducted since to enhance their rate capability - limited by the resistive electrodes [4].

A novel fast gas-avalanche detector concept was introduced [5], combining a Cherenkov radiator and a Micromegas (MM) multiplier. Photoelectrons (pe), emitted from a photocathode by particle-induced Cherenkov photons, are multiplied in a gaseous amplification structure. The authors proposed two modes of operation, incorporating semitransparent (ST) and reflective photocathodes, followed by a thin single-MM multiplication stage. The 1cm diameter detector prototype, with a ST CsI photocathode, yielded a time resolution of 27ps RMS for 50 detected photoelectrons per event. The idea of the mesh-coated with a reflective photocathode has not been followed, to the best of my knowledge, due to the very low effective-surface coverage – thus poor expected photoyield; different photoelectron paths into the multiplication gap also affecting time resolution. Further investigations have led to the two-stage PICOSEC concept [6]. The photocathode-coated radiator is followed by a two-stage MM multiplier; photoelectrons undergoing immediate preamplification in a very narrow (~100-200µm) gap, are transferred through a thin mesh into the similarly narrow MM stage. First results with a 1cm diameter detector yielded 24 ps RMS with MIPs [6]. An extensive R&D program, with dedicated 1cm$^2$ single-channel, two-stage MM PICOSEC detector prototypes, yielded respective time resolutions of 12.5, 32 and 34.5 ps with thin semitransparent CsI, Diamond Like Carbon (DLC) and $B_4C$ photocathodes deposited on a 3mm thick $MgF_2$ crystal radiator [7, 8]. The difference in time resolution results from the number of particle-induced photoelectrons (e.g. ~12 pe [6], ~2.5-3 pe and ~2-4 pe [7] for 18nm thick CsI, 1.5nm thick DLC and 9nm thick $B_4C$). The results were obtained at atmospheric-pressure $Ne/CF_4/C_2H_6$ (80/10/10). A recent work by the PICOSEC Micromegas collaboration [9] and references therein, summarizes the current MM-PICOSEC status, including that of larger, up to $20x20cm^2$, detectors and their related electronics. Note that in another configuration, a $1cm^2$ Microwell-PICOSEC with ST CsI photocathode provided ~24ps (RMS) resolution [10].

The intrinsic time resolution of the PICOSEC detector depends largely on the number of detected particle-induced photoelectrons. Their yield is derived from the Cherenkov-radiator thickness (naturally affecting, besides timing, their spatial distribution - thus event localization) and spectral transmission, the effective quantum efficiency (QE) of the photocathode and the detector's avalanche gain (signal-to-noise). The effective QE is dictated by photoelectron backscattering on gas molecules; thus, on the gas type and pressure and on the electric-field strength at the photocathode surface [11, 12]. The avalanche-multiplier's type and geometry, as well as its operation mode, electric-fields and other parameters, dictate the avalanche rapidity and size (gain); all, including the evident electronics' contribution - affect the time resolution.

While MM-PICOSEC has reached record time resolutions, by clever adjustment of its electrodes' configurations, low-capacitance "packaging" and low-noise electronics [9], some open questions persist regarding its applicability in "harsh environment" expected in future HEP experiments. On the one hand, radiation hardness of the radiator; on the other one, the relatively low QE and fragility of the ultrathin ST photocathode deposited on the radiator's surface. Fragility in terms of air exposure, avalanche-ion impact, eventual discharges, UV radiation and particle impact. Examples of thin CsI photocathode aging by air exposure are given in [13]. Examples of aging results of CsI photocathodes due to UV-photons' and avalanche-ions' impact are provided in [11, 14, 15]; that of thin DLC films, compared to CsI are depicted in [16]. While 50% drop in QE was reported for a 500nm thick reflective CsI photocathode, for an accumulated charge of ~ $14mC/cm^2$ [14], similar drop was observed for a 18nm thick ST one after only ~$0.3mC/cm^2$ [7].

The ST photocathode thickness is dictated, on the one hand, by the impinging UV-photon absorption length, and on the other by the photoelectron escape length from the photocathode. Both parameters were studied for thin CsI films, yielding values of the order of 20nm at 170nm wavelength [17, 18]. QE values of ST CsI were experimentally investigated by [19] and modelled in [20] as function of the film thickness. Optimal values of ~ 16% were reached for CsI thickness of ~20nm at 170nm photon wavelength – as compared to ~30% for 500nm thick ones [11]. For comparison, boron-doped, hydrogenated, heat-treated "thick" diamond films (with negative electron affinity, NEA) yielded at best QE values of ~6% [21]. While optimal ST photocathode thicknesses of 18, 9 and 3nm, were reported for respective CsI, $B_4C$ and DLC films [7, 8, 9], preliminary measurements reported by the PICOSEC Collaboration (yet unpublished), suggest higher PE yield values with few nm thick CsI films on Ti-coated $MgF_2$. The DLC and $B_4C$ photocathodes, of lower QE values, yielded lower photoelectron yields – reflected in the worse resolution times quoted above. Note that recent studies of powder-made NEA-nanodiamond films (thickness unknown) yielded at best QE values of ~1.5% at 170nm [22].

We propose here other detector concepts, combining electron multipliers with thick reflective photocathodes. The aims being enhanced photocathode robustness and quantum efficiency – thus expected good time resolutions due to larger Cherenkov radiation induced photoelectron yields. The detectors are defined here as Reflective PICOSEC, or R-PICOSEC.

## 2. Atmospheric R-PICOSEC

A potential reversed mode of operation of a PICOSEC detector at atmospheric pressure is depicted in Figure 1. In this configuration, a thick reflective photocathode is deposited on the readout pads (like in CsI-RICH detectors [23]). Note that the internal reflection losses of Cherenkov UV-photons at the crystal-gas interface depend on the particle's impact angle, affecting the emitted-photons yield – particularly close-to-normal incidence (see discussion below). Photoelectrons are preamplified in a first narrow gas gap; these initial-avalanche electrons are transferred into the second MM-like narrow parallel amplification gap through a resistive grid of high optical transparency (woven mesh or thin parallel wires), where multiplication occurs towards an ultrathin metal grid (of high UV transparency) deposited on the Cherenkov radiator (in a parallel-plate multiplication mode); the metal grid can be replaced by a microstrip configuration (a microstrip plate, MSP, with alternating anode and cathode strips), similar to that of Oed's Microstrip Gas Chamber (MSGC [24]). A lower field across the preamplification gap would assure good avalanche-electron transfer into the second stage and only partial ion-backflow (IBF) to the photocathode – advantageous for its long-term stability. In the MSGC mode, an additional fraction of the avalanche ions created near to the anode strips, are collected at the neighbouring cathode ones (Figure 2). This further reduces the ion backflow fraction to the photocathode.

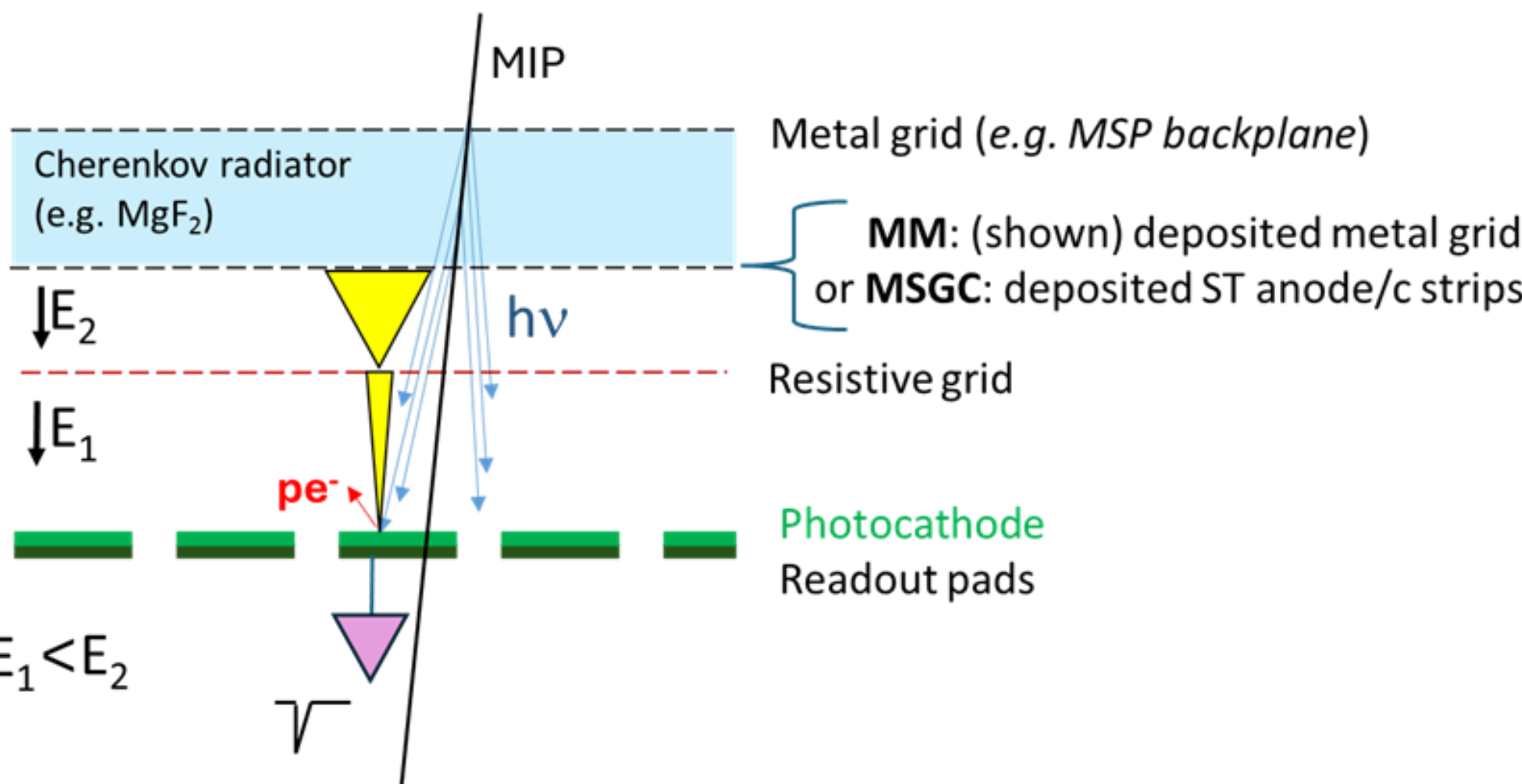

Figure 1. A reversed 2-stage mode of operation of a Reflective-PICOSEC detector. Particle-induced Cherenkov photons emit photoelectrons from a thick reflective photocathode deposited on readout pads. They are preamplified in a narrow gap and transferred to a second stage through a resistive grid. Amplification occurs in a MM mode (shown) towards a thin metal grid deposited on the crystal, or on deposited narrow anode strips in MSGC mode (not shown). Here, the charge induced by the avalanche, through the resistive grid, is detected on the pads. A large fraction of the second-avalanche ions is blocked by the grid. In the MSGC mode, some of the ions are additionally collected by nearby cathode strips (Figure 2).

The avalanche-induced charge propagates through the resistive grid onto the cathode pads (or strips); it provides time and position information. The few-hundred microns wide gap between the anode (or strips) deposited on the radiator and the photocathode-coated readout pads will keep the induced-charge on the latter – rather narrow.

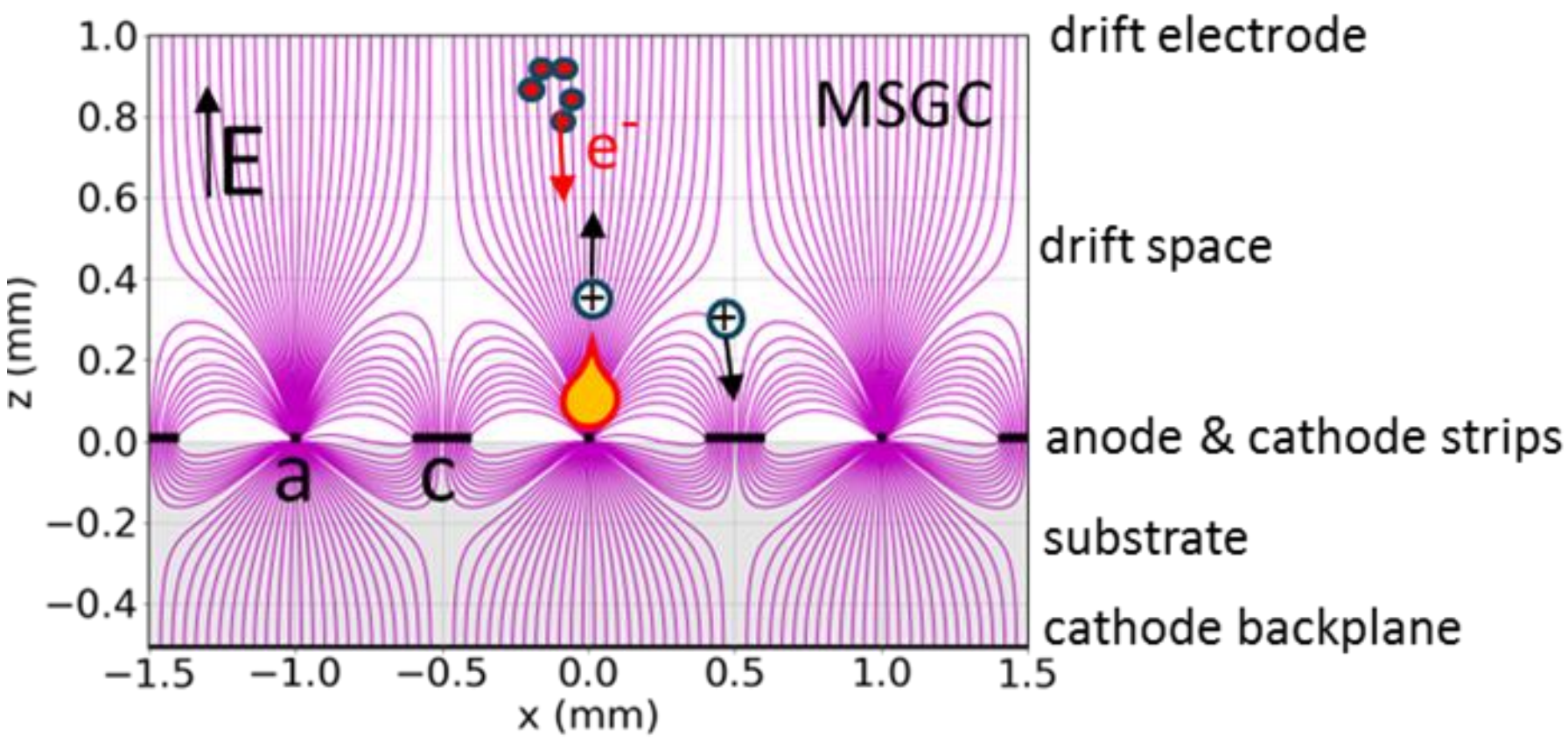

Figure 2. An example of field-line distribution in a MSGC configuration with anode and cathode strips deposited on an insulating substrate (e.g. the Cherenkov radiator crystal). It is dictated by the electrodes and strips' geometry and by the potentials applied to the various electrodes. Ionization electrons are multiplied at the vicinity of few µm wide anode strips; A fraction of the avalanche ions is collected on the nearby strips

The thin resistive grid may pose a certain challenge. In addition to the efficient transmission of the fast (sub-ns) [6] avalanche-induced signals, it should be mechanically stable and of high optical transmission. Woven meshes could be either made of a high-resistivity material or a polymer coated by a resistive material. Potential candidates could be Ge (Ge-oxide) coated polymer [25, 26], resistive polymer (e.g. resistive Kapton), etched Si etc. A resistive-wire mesh solution could be simpler. Properties of signal induction through resistive films can be found in [27]. The conductive anode grid or strips, vacuum-deposited on the radiator, should also have good optical transmission; an appropriate geometry and few nm thick Ni, NiCr or another UV-transmissive metal [28].

In parallel to vacuum deposition, other techniques could be considered. Besides CsI (CuI and NaI [29] as well as CsBr [14] were also investigated), thick diamond and $B_4C$ photocathodes mentioned above – other candidate photo-emissive materials deserve "engineering" and investigations. The accent should be on their QE, stability in short exposure to potential impurities (e.g. moisture) during transfer and mounting, radiation hardness, and long-term stability under exposure to Cherenkov UV-photons and avalanche ions.

## 3. Low-pressure R-PICOSEC

A most interesting R-PICOSEC concept could be of a multiplier operating at very low gas pressures, coupled to the reflective photocathode.

Low-pressure gas-avalanche detectors have been developed for many years. Their principal, but not only field of application has been nuclear physics; more specifically, low-energy heavy-ion experiments. The low-pressure operation in the latter (typically of a few-to-tens of mbar) was imposed by the necessity of carrying out such experiments in high-vacuum installations, with detectors equipped with sub-micron to few-micron thin entrance windows (to limit energy-loss effects). Among most popular ion tracking and timing detectors were low-pressure drift chambers [30, 31], parallel-plate ones [32] and multiwire proportional chambers (MWPC) [33]. Low-pressure operation of other multipliers, e.g. Multistep Avalanche Chambers (MSAC) [34], Microstrip Gas Chambers (MSGC) [35, 36], Microdot chambers [37], Gaseous Electron Multipliers (GEM) [38], Micromegas (MM) [39] and Thick GEM (THGEM) [40] has also been investigated in some other applications; among them in experiments requiring single-electron sensitivities [41, 42].

### 3.1 Low-pressure Multiwire R-PICOSEC

The first candidate could be a MWPC operating at a few mbar pressures.

Charpak's MWPCs operating at atmospheric pressure [43] have typical resolution times in the tens of ns range; it is derived from the long ionization-electron collection times in the gas volume prior to avalanche multiplication at the wire vicinity [44]. In contrary, it has been proven that the operation mechanism of MWPC at few mbar is different [45]. Under the intense reduced electric field (E/p) values, reaching >$10^3$ V/cm.mbar (at 1mbar isobutane) also at the so-called collection region, charge multiplication is characterized by a fast dual-step avalanche process. It is initiated at the location of the ionization-electron deposition and is followed by a fast-growing avalanche at the wires' vicinity (under fields >$10^5$ V/cm.mbar). Therefore, even with the estimated few ionization electrons deposited per mm of gas at 1mbar, at the vicinity of the MWPC cathode, intrinsic time resolutions of the order of 40ps (RMS) were reached with highly ionizing 27 MeV oxygen ions, with current pulses having ~2ns rise-time. The 3x3 $cm^2$ MWPC had 10μm diameter wires, 1mm apart, with 1.6mm anode-cathode gap; at 0.3-1mbar isobutane the charge gain was ~$10^5$ [33]. The low charge density under low-pressure avalanche and the fast ion removal, resulted in high-rate capability, e.g. a few % drop in pulse height was observed under heavy-ion irradiation of ~$5.10^5$ counts/s.$mm^2$ [45]. Larger-area (28x28$cm^2$) low-pressure MWPCs, deployed for the search of "strange matter" [46], reached resolution times of a few hundred ps (RMS) at 3mbar of isobutane; this was adequate for these experiments, with no electronics' precautions [47]. Low-pressure MWPC multipliers, equipped with a photocathode, have attained charge gains over $10^6$, when operated with single UV photons [45]; time resolution was not measured though in these conditions. It is interesting to add that localization resolutions of the order of 40μm RMS were reached with heavy ions at mbar pressures [48].

Two configurations are proposed for a Low-Pressure Multiwire (LPMW) R-PICOSEC, depicted in Figure 3. As alternative to the standard MWPC one (Figure 3a), the main goal being to attain both high charge gains and low avalanche-ion backflow (IBF). This can be reached, on the one hand, by inserting a thin resistive grid (Figure 3b) between the photocathode and the wires, with fields arranged to pre-amplify and efficiently transfer the avalanche electrons, while blocking part of the back-drifting ions. On the other hand, a large fraction of ions can be collected by deploying alternating anode and cathode wires (Figure 3c) [49]. In both cases the wires are located at a short distance from the radiator's surface; the latter being coated with a thin UV-transmissive metal grid field-shaping cathode. In the first configuration, timing and localization are performed with charge-induced signals, propagating through the high-resistivity grid, towards the photocathode-coated pads or strips. A resistive grid of a few Megaohm/square would transmit most of the induced charge [27], including fast current signals [26]. In the alternating-wires configuration, it is expected that a large fraction of ions drifting rapidly away from the anode wires will induce fast signals on the photocathode pads. Note, that the avalanche is initiated at the photocathode vicinity; therefore, initial-avalanche ions, created below the resistive grid or the wires' array will reach the photocathode – though with minimal contribution to the signal and to ion-impact aging.

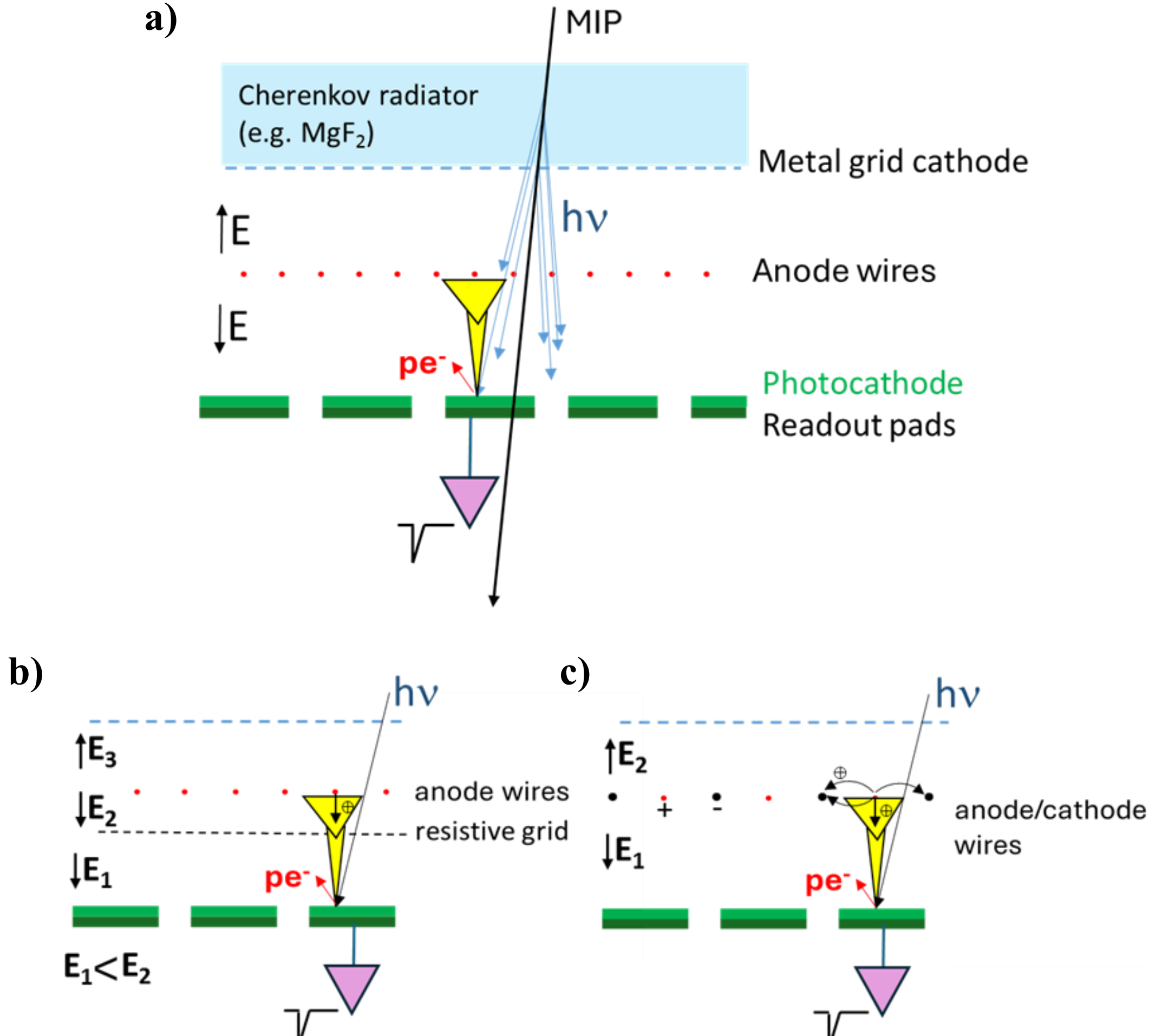

Figure 3 a) A reversed operation mode of a low-pressure MWPC-based Reflective-PICOSEC detector. Photoelectron avalanche starts at the photocathode vicinity under very high reduced electric field E1/p values; the final avalanche occurs at the thin wires. The avalanche-induced signals are detected on the photocathode-coated pads. Avalanche-ions backflow to the photocathode can be reduced by b) the insertion of a resistive grid between wires and pad, transparent to the induced charge; c) by deploying alternate anode and ion-collecting cathode wires. Note that a resistive grid might be inserted also under the alternating wires – to enhance ion blocking.

### 3.2 Low-pressure Microstrip R-PICOSEC

A more compact low-pressure R-PICOSEC concept would be that of a reflective photocathode coupled to a microstrip electrode [24] deposited on the Cherenkov radiator, in similar configurations discussed above. A Low-Pressure Microstrip (LPMS) R-PICOSEC detector is depicted in Figure 4. Unlike in the LPMW, the final charge multiplication in the LPMS, that follows gap preamplification under the high reduced electric field values [35], occurs on strips deposited directly on the Cherenkov-radiator’s exit surface. A thin cathode grid, the backplane electrode, is deposited on the radiator’s input face.

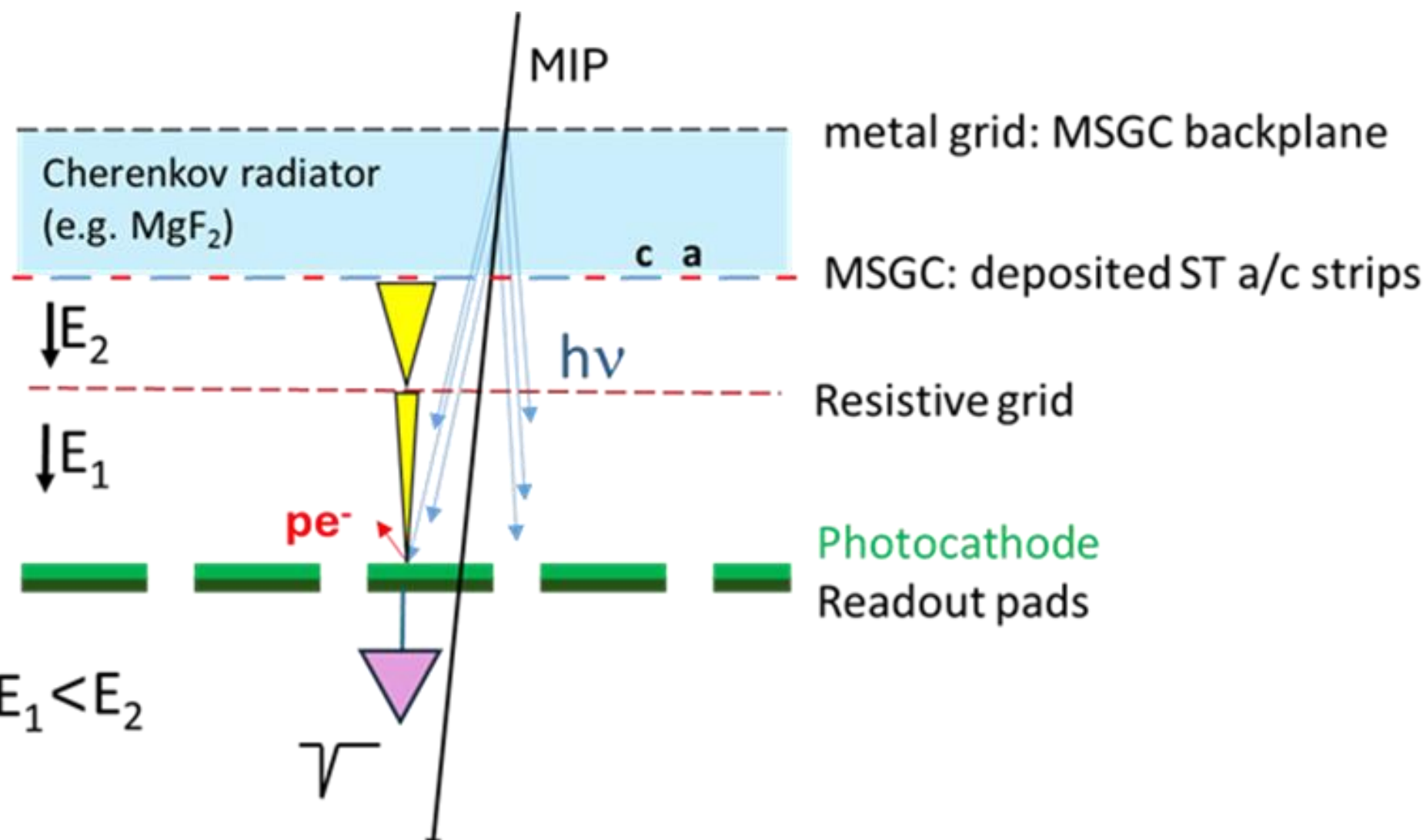

Figure 4 A reversed operation mode of a low-pressure Microstrip (LPMS) Reflective-PICOSEC detector. Photoelectron avalanche starts at the photocathode vicinity under very high reduced electric field E1/p values; the avalanche electrons are transferred through a resistive grid to a second parallel gap, where the final avalanche occurs, mostly on the MSP anode strips. Avalanche-ions backflow to the photocathode is largely reduced by their collection at the cathode strips (Fig.2) and blocking the remaining ones by the grid. The avalanche-induced signals are detected on the photocathode-coated readout pads.

While here we refer to the “classical” MSGC configuration, with alternating anode and cathode strips [24], other configurations could be considered as well [50, 51, 52]. While designing the detector to maximize the charge gain, care should be taken to trap a large fraction of the avalanche ions. As discussed above, an important advantage of the MSGC geometry is the rapid side-collection, to the nearby cathode strips, of a large fraction of the ions generated in the anode-strip avalanche – considerably reducing ion backflow to the photocathode. In addition, one could insert also in the LPMS operation mode, a resistive mesh between the Microstrip electrode and the photocathode-coated readout pads. Electric-fields optimization would further reduce the avalanche-ions’ backflow to the photocathode.

The LPMS operation was validated with a $10cm^2$ MSGC detector equipped with a photocathode coupled to the anode-strips’ electrode [35]. With 8µm anode strips, 200µm apart, interlaced by 80µm wide cathode ones, the MSGC yielded charge gains of the order of $5x10^4$ at 25mbar of Isobutane. The time resolution was not measured but the fast current pulses had a few ns rise-time with single photoelectrons, at these operation conditions. Similar fast two-stage mode of operation and results were reached in a Microdot detector operated at a few tens of mbar [37].

It should be reminded though, that Microstrip and other multiplying structures with charge multiplication occurring on metallic electrodes deposited on insulating substrates are prone to charging up and surface polarization phenomena. This requires precautions in the selection of detector design and substrate (here the radiator crystal) materials and their eventual surface treatments as discussed in [53, 54].

Note, that the total charge gain, avalanche rapidity, ion blocking and time resolution will depend on the detector parameters, gas and pressure. The time resolution will naturally largely depend also on the photocathode properties – dictating the photoelectron yield - as discussed above.

## 4. Summary and Discussion

The ideas presented in this work aim at conceiving alternative PICOSEC configurations, with the potential of overcoming some drawbacks of current detector concepts. Unlike "standard" PICOSEC designs, with few-nm thick semitransparent (ST) photocathodes deposited on the Cherenkov radiator, we propose various other configurations and operation modes. They incorporate thick reflective photocathodes deposited on the readout electrodes of various types of avalanche multipliers - operating at atmospheric and few-mbar gas pressures. The main goal being the enhancement of the photoelectron yield, charge gain, avalanche development rapidity – all affecting the time resolution; an important aim being also the enhancement of the detector robustness.

The concepts discussed, are naturally immature; they require further thinking, evaluation of technologies involved, simulations and experimental investigations. While physics-wise, they might be sound, many pitfalls exist.

One of the first questions coming to mind relates to the photoelectron yield resulting from a relativistic particle interacting with the radiator crystal; namely the difference between the current situation of a thin ST photocathode deposited on the crystal (on top of a few nm thin conductive layer), and a thick reflective photocathode located in the gas volume (or vacuum) in front of the crystal.

Figure 5 depicts preliminary estimation results of photoelectron yields for 100GeV/c muons traversing a 3mm thick $MgF_2$ radiator, vs the muon's angle of incidence. Two configurations: a 18nm ST CsI film [7] is deposited on top of a 2nm thick Cr film coating the crystal, and a 500nm thick reflective CsI photocathode being placed in vacuum in front of the same crystal. CsI QE values were taken from [55].

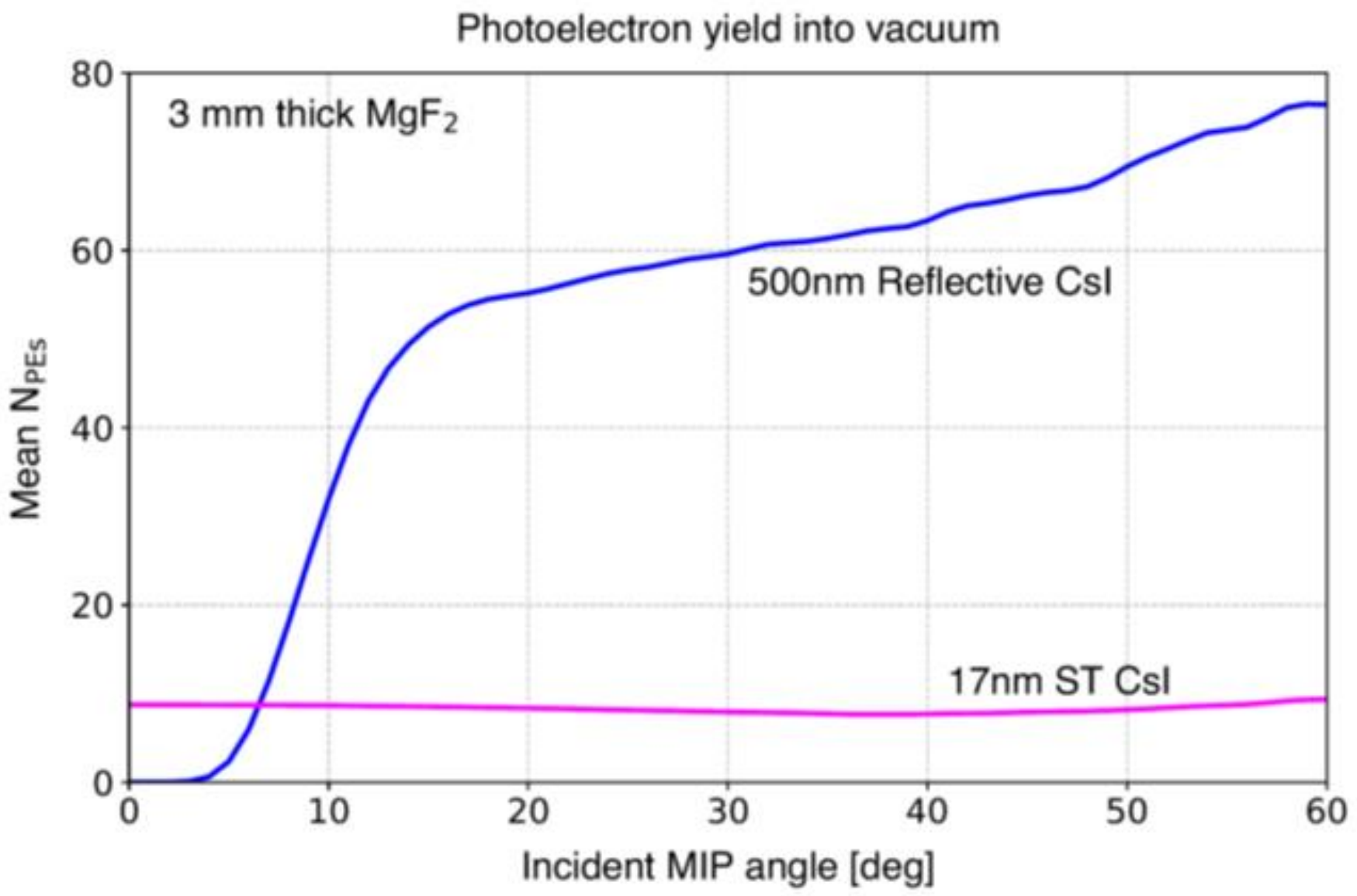

Figure 5. Estimated photoelectron yields emitted into vacuum, from a semitransparent CsI photocathode deposited on a Cr-coated 3mm thick $MgF_2$ Cherenkov radiator and that from a thick reflective one facing the bare crystal. The photoelectron-yield curves are plotted vs the relativistic-particle's incidence angle to the normal (0 deg is normal incidence). Cherenkov photon spectral range: 120-200nm QE data from [55].

Literature provides a broad range of data for the QE of reflective CsI photocathodes, of which that of [56] covers the spectral range of 125-220 nm. Data for ST photocathodes are scarce, varying according to the photocathode production and transfer – affecting their surface morphology. Therefore, the data of [55] were chosen – combining both, ST and reflective CsI measured presumably in similar conditions. The index of refraction vs wavelength for $MgF_2$ was taken from [57]. As reflected from figure 5, above ~5 deg, the photoelectron yield of the reflective photocathode configuration surpasses that of the ST one. At lower angles, photon emission from the crystal to vacuum is affected by the critical angle – suggesting a slightly tilted detector deployment. More elaborate calculations are necessary.

Multipliers with metal grids or strips deposited on an insulator, as proposed in this work (on the Cherenkov-radiator crystal) are prone to instabilities – originating from surface charging up and surface polarization phenomena. This requires precautions in materials selection and multiplier design, discussed above. Unlike substrate coating with ultrathin resistive films, or ion implantation, that might absorb a certain fraction of the Cherenkov UV photons (as discussed, in addition to other techniques in [53]) should be investigated. One could also consider replacing the currently employed $MgF_2$ Cherenkov radiator with other crystals of known lower resistivity values - e.g. LiF. Note that in gas media, according to impurities, the literature-quoted surface-resistivity (or sheet-resistivity) values might not provide realistic data; thus, necessitating dedicated studies.

Some of the concepts above, relying on avalanche induced-charge transfer onto readout pads (e.g. Figures 1, 4), involve the deployment of thin ion-blocking resistive grids (wires or woven mesh). In addition to the investigations of appropriate resistive materials, stretching such thin grids while keeping narrow multiplication gaps requires feasibility studies. Another concern might be timing properties relying on charges induced on the cathode pads through a resistive grid. Parameters influencing the resulting pulse amplitude and shape, besides the avalanche-buildup and charge transport properties, are the grid resistivity and multiplier geometry – affecting the RC time constant of the structure. Though the avalanche being initiated at the photocathode surface, the timing properties of reversed configurations should be carefully investigated, in comparison to that of the direct signal readout on anode pads. Note that cathode-induced signals in cascaded DLC-coated resistive-GEM configurations were proposed for Fast Timing Detectors (FTM) [58]; however, the authors do not provide information on the current-pulse attenuation and shape. Similar DLC-coated perforated GEM-like foils, with large optical transparency, would also deserve investigations as "resistive grids". Other properties, related to the broader distribution of the charge induced on the pad electrodes, also require consideration, e.g. their effect on the localization of the Cherenkov-photons' patterns.

The idea of avalanche-ion "side-collection" by neighbouring cathode wires/strips in multiwire and in microstrip multipliers requires careful simulations and investigations – with the aim of maximizing ion trapping. Some studies were carried out in low-pressure MSGC detectors [59]. Figure 6 depicts the 2-stage detector geometry and results of charge separtition between the detector's electrodes at 13 mbar of isobutane. In this example, under the indicated conditions, in the two-stage avalanche mode (gap gain~100; strips gain ~10) only 40% of the ions reach the photocathode in the optimal condition. The ion trapping is expected to be far more significant with varying the gain contributions of the two elements and inserting the blocking grid, as discussed above.

Ultrafast avalanche formation at low gas pressures, validated long ago with multiwire and microstrip detectors with heavily ionizing particles, could be a good way to go. While high charge gains for single electrons were reached at mbar isobutane pressures with MWPCs, microstrip detectors were investigated only down to 15mbar. No photon-feedback effects were observed with the presence of CsI photocathodes. Unlike "standard" atmospheric-pressure operation, both detector configurations operated with charge preamplification within the "collection region", due to the very high reduced fields (E/p) values; it was followed by the final multiplication near thin wires or strips. The MWPC provided ~40ps RMS time resolutions, with few ion-induced electrons per mm, at 1mbar isobutane. Thus, good timing properties are expected for surface emitted photoelectrons, initiating avalanche at their emission location - provided that the wire or strip geometry in the final stage does not affect considerably the avalanche-electrons paths. The low-pressure operation would naturally require careful mechanical design of the detectors.

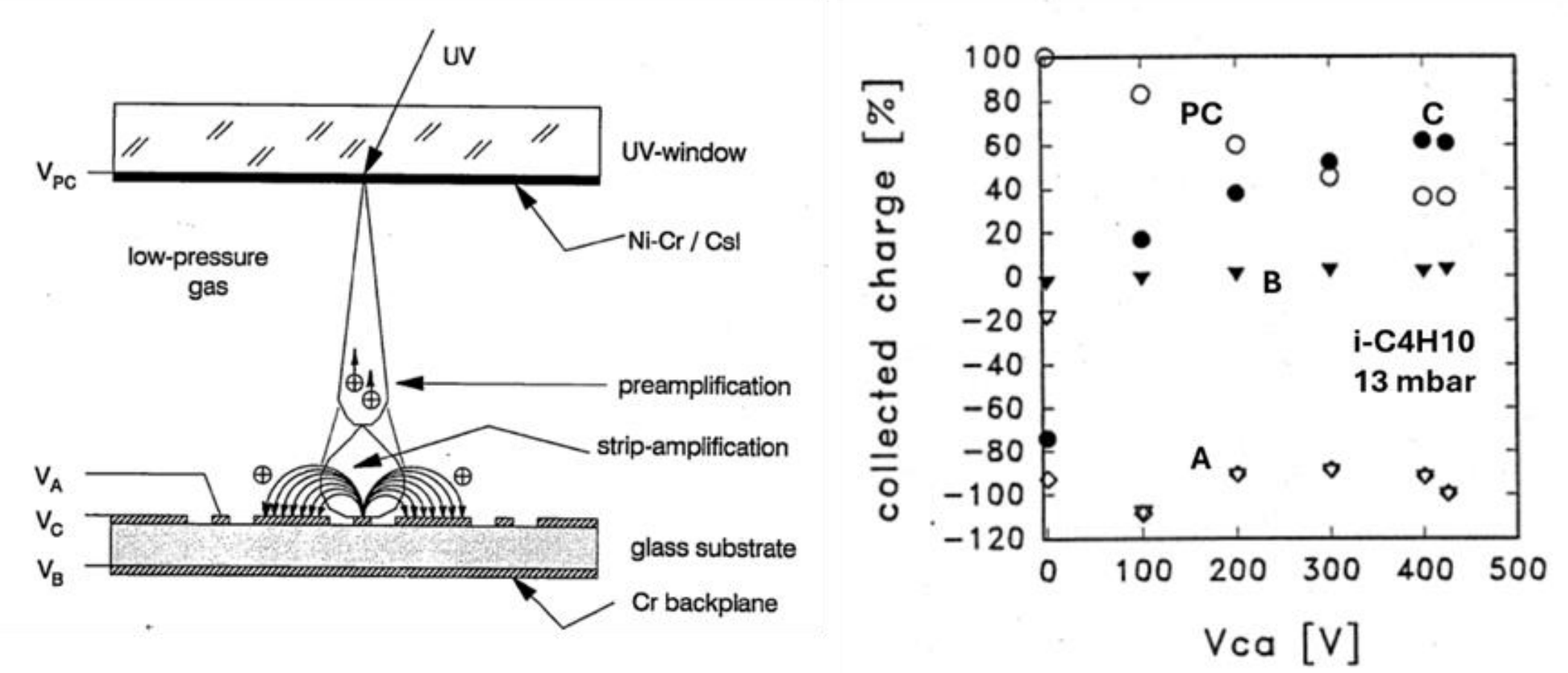

Figure 6. Left: A UV-photon detector combining a semitransparent CsI photocathode coupled to a low-pressure MSGC electron multiplier. Right: Fraction of collected avalanche-induced charge on the different electrodes of the shown detector - vs voltage difference between anode and cathode strips - $V_{CA}$. PC-photocathode; C-cathode strips; B-backplane; A-anode strips. Detector operated at 13mbar isobutane. $V_{PC}$=-870V; $V_B$=$V_C$= -100V. Single photoelectron gain ~$10^5$. Figures taken from [59].

In the search for robust, highly emissive photocathodes, one should also consider their charging up effects at high counting rates, due to their relatively high bulk resistivity. Discussion on this matter and results for reflective CsI photocathodes are provided in [15]. Another relevant topic would be the extension of the spectral response towards the visible range. It is known, however, that the lifetime of visible-sensitive photocathodes (e.g. $Cs_3Sb$, $K_2CsSb$, GaAs etc.) is affected even by minute amounts of impurities in vacuum and more dramatically – in gas media. The short-term operation of a gaseous photomultiplier with a bare $K_2CsSb$ photocathode coupled to a cascaded Gas Electron Multiplier and Micro-hole and Strip Plate (GEM-MHSP) multiplier, was demonstrated though in [60]. Extensive studies have been carried out to coat visible-sensitive photocathodes with thin protective films, e.g. CsI and others. They resulted in gain of their stability but loss of QE in the visible [61, 62, 63]. Interestingly though, the spectral sensitivity of these "composite photocathodes" ranges from the far-UV to the visible. Another idea was to coat the photocathode with a removable organic film [64], just for its safe transfer into a detector. Note that considerable progress has been made since in the field of photocathode-surface protection with Graphene and other 2D atomic layers – mainly for laser-driven electron sources [65, 66]. This could suggest similar R&D in the photon-detectors field.

## Acknowledgments

I would like to dedicate this article to my friend and closest collaborator for many decades, Rachel Chechik, on her 80th anniversary. I am indebted to her and to many other close colleagues and generations of students and young researchers who accompanied me along my long Detector-Physics path. Many of the ideas presented here are based on our common research results.

I warmly thank Darina Zavazieva and Roy Gil of Technion for their assistance with the simulations and members of the PICOSEC collaboration for useful discussions; I appreciate the kind assistance of Fabio Sauli, Leszek Ropelewski, Eraldo Oliveri, David Vartsky, Shikma Bressler and Lucian-Simon Scharenberg in providing useful comments to this work.